\documentclass[10pt,conference]{IEEEtran}
\ifCLASSINFOpdf
\else
\fi
\let\labelindent\relax
\usepackage{hyperref}
\usepackage{graphicx}
\usepackage{adjustbox}
\usepackage[utf8]{inputenc}
\usepackage{booktabs}
\usepackage{subfigure}
\usepackage{tcolorbox}
\usepackage{amsmath}
\usepackage{amssymb}
\usepackage[numbers]{natbib}
\usepackage{paralist}
\usepackage{multirow}
\usepackage{xspace}
\usepackage{color}
\usepackage{xcolor}
\usepackage[graphicx]{realboxes}
\usepackage{ifthen}
\usepackage{url}
\usepackage{fancybox}
\usepackage{enumitem}
\usepackage{listings}
\usepackage{balance}
\usepackage{ifthen}
\usepackage{graphicx}
\usepackage{amsmath}
\usepackage[linesnumbered,boxruled]{algorithm2e}
\usepackage{algorithm2e}
\usepackage{tablefootnote}
\usepackage{comment}
\usepackage{float}
\usepackage{algorithmic}
\usepackage{dirtytalk}
\usepackage{subfigure}
\usepackage[tikz]{bclogo}
\usepackage[colorinlistoftodos]{todonotes}
\usepackage{epigraph}
\usepackage{booktabs}
\usepackage{tikz}
\usepackage{tcolorbox}
\usetikzlibrary{plotmarks}
\usetikzlibrary{arrows,shapes,positioning}
\usetikzlibrary{decorations.markings}
\tikzstyle arrowstyle=[scale=1]
\tikzset{>=latex}

\usepackage{pgfplotstable}
\graphicspath{{pics/}}

\definecolor{codegreen}{rgb}{0,0.6,0}
\definecolor{codegray}{rgb}{0.5,0.5,0.5}
\definecolor{codepurple}{rgb}{0.58,0,0.82}
\definecolor{backcolour}{rgb}{0.95,0.95,0.92}

\lstdefinestyle{mystyle}{
	backgroundcolor=\color{backcolour},   
	commentstyle=\color{codegreen},
	keywordstyle=\color{magenta},
	numberstyle=\tiny\color{codegray},
	stringstyle=\color{codepurple},
	basicstyle=\footnotesize,
	breakatwhitespace=false,         
	breaklines=true,                 
	captionpos=b,                    
	keepspaces=true,                 
	numbers=left,                    
	numbersep=2pt,                  
	showspaces=false,                
	showstringspaces=false,
	showtabs=false,                  
	tabsize=2
}

\usepackage[english]{babel}

\DeclareGraphicsExtensions{.pdf,.jpeg,.png}
\begin{document}

\newtheorem{theorem}{Definition}[section]
	\newcommand{\eg}{e.g.,}
	\newcommand{\ie}{i.e.,}	
	\renewcommand{\lstlistingname}{Listing}

\newcommand{\boxedtext}[1]{\fbox{\scriptsize\bfseries\textsf{#1}}}
\newcommand{\nota}[2]{
	\boxedtext{#1}
		{\small$\blacktriangleright$\emph{\textsl{#2}}$\blacktriangleleft$}
}

\newcommand\autcomment[1]{{\textcolor{pink}{\textbf{}#1}}}

\newcommand\ct[1]{\textcolor{pink}{\textbf{Christoph:}} {\textcolor{pink}{#1}}}

\newcommand\raula[1]{\textcolor{red}{\textbf{Raula:}} {\textcolor{blue}{#1}}}

\newcommand\asia[1]{\textcolor{red}{\textbf{Asia:}} {\textcolor{red}{#1}}}

\newcommand\review[3]{\textcolor{red}{\sout{#1}} {\textcolor{blue}{#2}}{\todo{#3}}}

\newcommand{\Rqone}{RQ$_1$: \textit{To what extent do developers make contributions to the ecosystem?}}
\newcommand{\Rqtwo}{RQ$_2$: \textit{What kinds of contributions are made to the ecosystem?}}
\newcommand{\Rqthree}{RQ$_3$: \textit{What are the motivation to contribute to the  ecosystem?}}

\definecolor{beaublue}{rgb}{0.74, 0.83, 0.9} 
\definecolor{redbeau}{HTML}{cc5b5b}  
\definecolor{orabeau}{HTML}{ff9f79}

  \title{We Live in a Society: \\ Motivators for Contributions in an OSS Ecosystem}

\author{Supatsara Wattanakriengkrai, Raula Gaikovina Kula, Christoph Treude, Kenichi Matsumoto \\
Nara Institute of Science and Technology, Japan, University of Melbourne, Australia\\
Email: \{wattanakri.supatsara.ws3,raula-k,  matumoto\}@is.naist.jp, christoph.treude@unimelb.edu.au}
\maketitle
\begin{abstract}
Due to the increasing number of attacks targeting open source library ecosystems, assisting maintainers has become a top priority.
This is especially important since maintainers are usually overworked.
Although the motivation of Open Source developers has been widely studied, the extent to which maintainers assist libraries that they depend on is unknown.
Surveying NPM developers, our early results indicate a difference in motivation between maintaining their own library (i.e., more person driven), as opposed to professional factors (i.e., focus on skills and expertise) when contributing to the software ecosystem.
Finally, our thematic analysis shows different motivations and barriers developers face when contributing to the ecosystem. 
These results show that developers have different motivations and barriers depending on the role they play when making contributions to the ecosystem.
\end{abstract}


\epigraph{\textit{``OSS can be described as a progression through 3 roles. Author, Contributor, Maintainer. The responsibilities, motivations, etc change drastically. A lot of emphasis is placed on authoring and contributing to OSS but very little is placed on the Maintainers that keep the entire OSS ecosystem alive and evolving.''}}{Participant 19}

\section{Introduction}
\label{sec:intro}

Third-party libraries are vital for modern software applications, not only for open source, but also in the industrial setting \cite{Vargas:fse2020}.
However, growing threats of instability have raised concerns in terms of the workload of library maintainers. 
For example, Google and Microsoft recently announced initiatives, such as the Alpha-Omega Project, to improve global OSS supply chain security.
They described \textit{``Alpha will work with the maintainers of the most critical open source projects to help them identify and fix security vulnerabilities, and improve their security posture.  Omega will identify at least 10,000 widely deployed OSS projects where it can apply automated security analysis, scoring, and remediation guidance to their open source maintainer communities''} \cite{AlphaOme74:online}.
Instability can also be caused by maintainer frustration (log4Shell) \cite{YfryTchs40:online} and even hijacking (faker, colors, rc) \cite{npmLibra85:online}, involving sabotage by the maintainer.

Most typical ecosystems are known to act like societies, maintainers of a library should receive and submit external contributions within the ecosystem, as they are perceived to share similar goals and motivations. 
Attracting such external contributions ensures a healthy flow of contributions (i.e., fix bug, add new features). 

Since most third-party libraries rely on volunteer contributions (usually unpaid and overworked), the abandonment of these contributors poses potential risks to the sustainability of these libraries \cite{2017_ruby}.
For example, Valiev et al. \cite{fse2018_sustained} showed that libraries with more contributors are less likely to become dormant.
Wattanakriengkrai et al. \cite{Supatsara:tse2022} also revealed that libraries indeed rely on the ecosystem for contributions, and these contributions share an inverse association with the likelihood of a library becoming dormant.
Furthermore, prior work has shown that the sustainability of libraries is not only affected by factors internal 
to the library, but also by the ecosystem \cite{Bogart2016breakAPI}, e.g., dependencies \cite{Golzadeh:2019}.
In particular, threats such as vulnerabilities \cite{chinthanet2021lags, Durumeric2014Heartbleed} and transitive dependency changes \cite{dey2019_promise_mockus} may cause a library to risk becoming
obsolete \cite{kula2018developers}.
Other prior work has found that these ecosystems form a complex web of dependencies, with each library dependent on each other.
Wittern et al. \cite{wittern2016look} analyzed a subset of NPM libraries, to suggest that NPM is a dynamic ecosystem with ongoing and even rapid growth of libraries and increasing dependencies between them.
Abdalkareem et al. \cite{abdalkareem2017developers} observed that NPM trivial libraries rely on a large number of libraries and depending on these trivial libraries can be useful and risk-free if they are well implemented and tested. 
 The dependence on other libraries in often brittle dependency chains implies that local sustainability issues around individual libraries can have widespread network consequences \cite{10.1109/MSR.2017.55}.
These studies are only concerned with the technical dependencies, and they do not focus on the social networks and human factors that exist within these ecosystems.

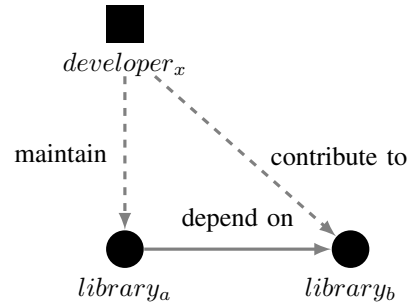
\begin{figure}[t]
    \centering
  
    \begin{tikzpicture}[
        roundnode/.style={circle, fill=black, minimum size=5mm},
        squarenode/.style={fill=black, text=red, minimum size=5mm},
    ]
    \begin{scope}
        
        \node[roundnode, label=below:$library_{b}$] (s2_projj) at (5,0) {};
        \node[roundnode, label=below:$library_{a}$] (s2_projk) at (2,0) {};
      
        \node[squarenode, label=below:$developer_{x}$] (s3_devx) at (2, 3) {};

    \end{scope}

    \begin{scope} [every edge/.style={draw=gray, very thick}]
     
        \path [->] (s2_projk) edge node[above = 1mm] {depend on} (s2_projj);
     
    \end{scope}
    \begin{scope} [every edge/.style={draw=gray, very thick}]
       
        \path [->][dashed]  (2, 2.3) edge node[left = 1mm] {maintain} (s2_projk);
        \path [->][dashed]  (2.4 , 2.3) edge node[right = 2mm] {contribute to} (s2_projj);
    
    \end{scope}

    \end{tikzpicture}
    
    \caption{Different types of contributions within an ecosystem}
    \label{fig:client-library}
\end{figure}

\begin{figure*}[t]
    \centering
    \includegraphics[trim=0 2 0 0,clip,width=0.9\linewidth]{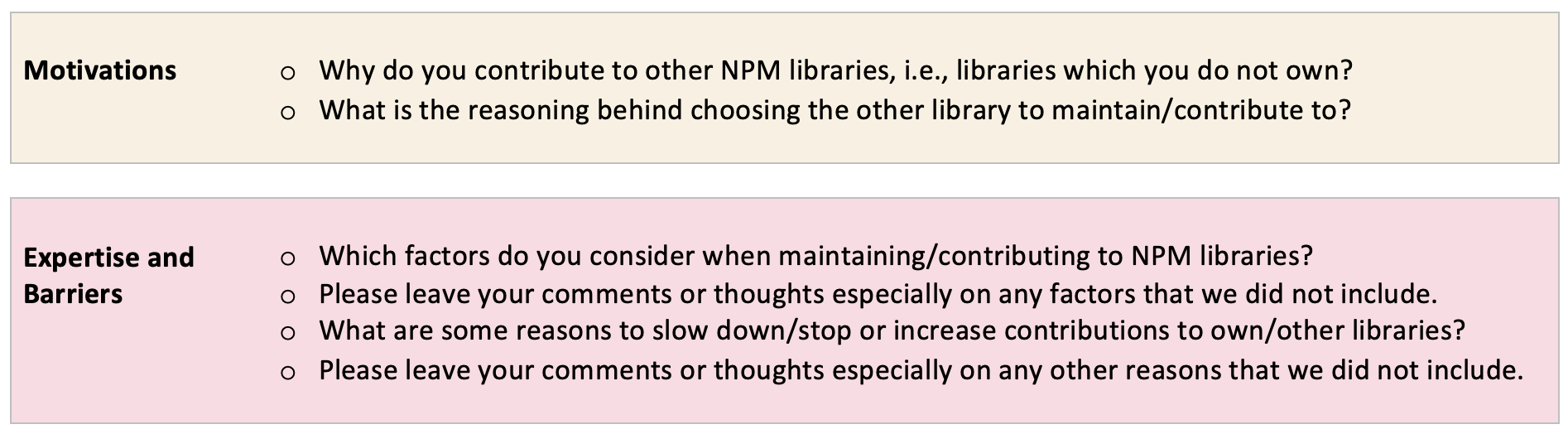}
    \caption{Qualitative survey questions divided into motivation, expertise and barriers}
    \label{fig:survey_question}
\end{figure*}

The new idea of this paper is to take a different perspective on understanding developer motivation, \textbf{on what motivates a developer contribution to the ecosystem (i.e., libraries that they depend on), as opposed to maintaining their own libraries}.
Figure \ref{fig:client-library} visualizes the client-library relationship, where there is a maintainer ($developer_{x}$) of a library ($library_{a}$) that is dependent on another library ($library_{b}$) and contributes to their dependency.  

We conducted a survey with NPM developers who host their projects on GitHub.
We selected the NPM ecosystem, as it is one of the largest ecosystems with more than two million libraries and has recently been under attack, as mentioned above. 
We conducted a survey with 49 participants who are maintainers of NPM libraries hosted on GitHub.
Using thematic analysis, our early results indicate four themes that changed the motivation of developers. 
Interestingly, we find that developers may show a difference in motivation when making a contribution to the OSS ecosystem, as opposed to contributing to their own libraries.

\section{Individual Motivation Viewpoints}
Existing work that studies the motivation of OSS project contributors takes a generic and individualist point of view.
For example, 
Lee et al. \cite{Lee:icse2017} investigated the motivations of one-time code contributors and found that developers primarily made contributions to fix bugs that hindered their work.
Similarly, Coelho et al. \cite{Coelho:chase2018} reported that the main reason core developers contribute to OSS projects is to improve a project because they are also using it.
A recent study by Gerosa et al. \cite{Gerosa2021TheSS} on the motivations of OSS developers provided evidence that intrinsic and internalized motivations, such as learning and intellectual stimulation, are highly relevant to many developers.

Expertise can be considered as a factor that drives developers to contribute to OSS projects.
For example, Vadlamani et al. \cite{Vadlamani:icsme2020} presented a qualitative study of the expertise of software developers to understand the factors that drive developers to contribute to OSS projects. 
They revealed that personal drivers (i.e., self-needs and hobby) are more critical than professional factors (i.e., skills and expertise) in motivating GitHub developers.

Recent work also acknowledges barriers to new core contributors.
For example, Avelino et al. \cite{Avelino:esem2019} investigated the abandonment of OSS projects by their main developers. The authors showed that projects survive such situations by attracting new core contributors, while there are various barriers that the new core developers faced when making contributions to these projects, such as lack of time and the difficulty to obtain push access to the repositories.

\begin{figure}[t]
\centerline{\includegraphics[trim=0 2 0 0,clip,width=1\linewidth]{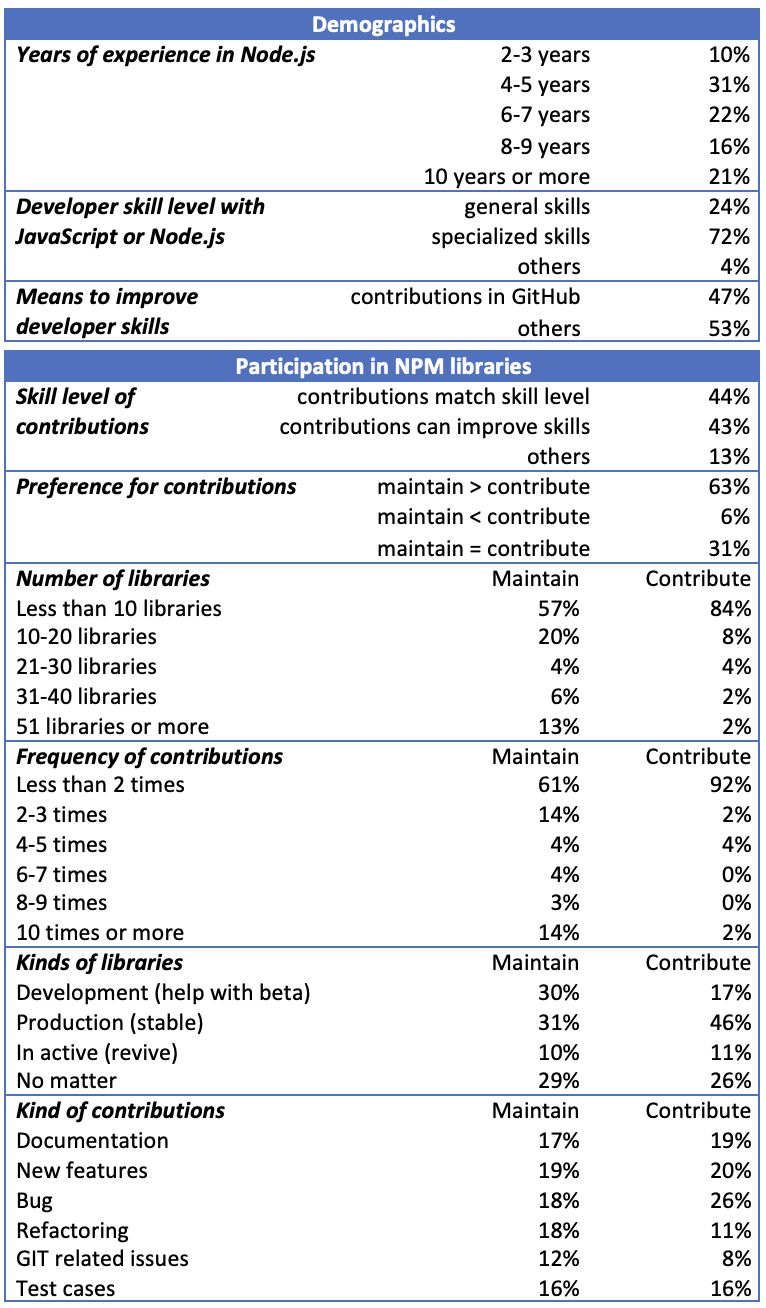}}
    \caption{Overview of responses from 49 Node.js developers.}
    \label{fig:demo_graph}
\end{figure}

\section{Research Method}
\label{sec:research_method}
In this section, we describe our survey design, recruitment, and analysis methods for the study.

\subsection{Survey Design and Recruitment}
 With Ethical Board approval\footnote{ethical approval documentation will be made available} from the anon institute, we contacted developers via email, with survey data available online\footnote{authors will make link available once accepted}.
Following previous work \cite{chinthanet2021lags}, we collected and sent out invitations via a curated list of developers.
Before distributing our survey to a wide range of participants, we conducted a pilot survey to confirm the understanding of the participants and their feedbacks allowed us to reformulate some questions.
The final version of the survey contained the following questions:
after asking about demographic information (years of experience, skill level, and whether they use contributions to improve skills) of the participants, we asked a set of contribution-related questions about participants' activities with their own libraries and with other libraries (preference for contributions, number of libraries, frequency of contributions, kinds of libraries, and kinds of contributions).

Figure \ref{fig:survey_question} details the qualitative questions we asked the participants. 
We divided the questions into two parts, allowing participants to also provide their comments and additional points of view. 
The first section is related to the motivation of a maintainer to contribute to other libraries. 
For the quantitative component, we added five choices (found a bug, depend on, new feature, issues of interest, and contributor familiarity).
These choices are based on Subramanian et al. \cite{Subramanian:IEEE2022}.
The second section was used to ask about the expertise and barriers that maintainers face when making contributions. 
The first question uses the same drivers as Vadlamani and Baysal \cite{Vadlamani:icsme2020}. 
As mentioned, personal factors are more self-needs and hobbies, while professional factors are more related to their specialized skills and expertise.
For the second question, we list four barriers (library stability, lack of time, motivation, and end of library lifecycle) based on Vadlamani and Baysal \cite{Vadlamani:icsme2020}. 

\subsection{Manual Coding}
To analyze responses to the qualitative questions (Figure \ref{fig:survey_question}), we used the card sorting method similar to previous work \citep{spencer2009card}, \citep{zimmermann2016card}.
The coding used and extended the intrinsic and extrinsic classifications proposed by Gerosa et al. \citep{Gerosa2021TheSS}.
The process is as follows.
First, the first two authors were assigned to conduct the card sorting task.
They reviewed all responses and created a ``card'' for each of them.
For validation, instead of coding the cards separately in parallel and checking the consistency of the coding results, the two authors coded the cards together \citep{begel2014analyze}, \citep{guzzi2013communication}.
Coders used an open-coding approach \citep{zimmermann2016card} to analyze the cards, where new themes emerged during the coding process.
The agreement was negotiated along the way \citep{campbell2013coding}, i.e., when the coders had different opinions, we interrupted the process to discuss the discrepancy before continuing. 

\begin{figure}[t]
    \centerline{\includegraphics[ width=0.9\linewidth]{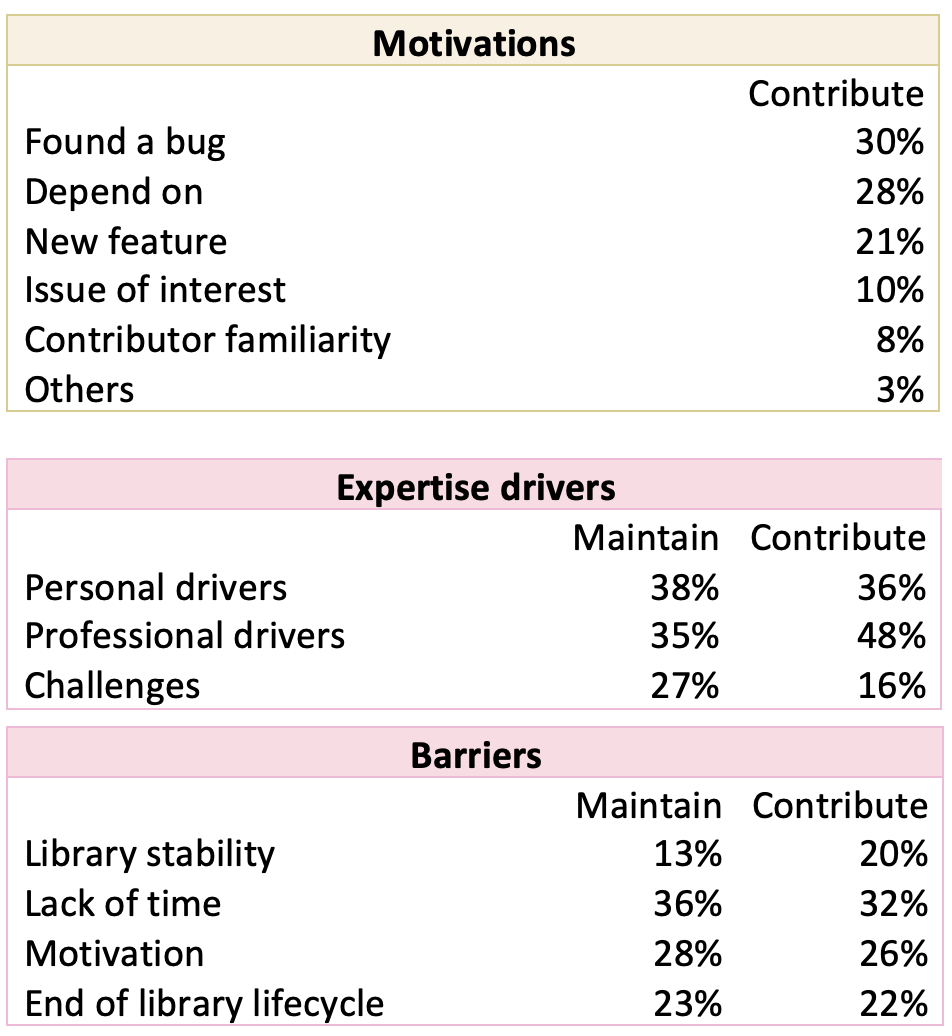}}
    \caption{Motivations, expertise, barriers for maintainers.}
    \label{fig:R11}
\end{figure}

\subsection{Respondent Demographics}

Figure \ref{fig:demo_graph} shows the demographics of the 49 participants.
We find that experience ranges from experts (ten years of experience or more -- 21\%), intermediate (four to nine years of experience -- 69\%, to beginner (two to three years of experience) -- 10\%).
Most participants reported having specialized skills. 
Interestingly, participants did not particularly use their contributions to improve their skill levels. 

\begin{table*}[ht]
\centering
\caption{Themes from our coding}
\label{tab:my-table}
\begin{tabular}{@{}lllr@{}}
\toprule
 &  Themes  & Participants & \# of responses \\ \midrule
&Own-use-client & \begin{tabular}[c]{@{}l@{}}P2, P7, P8, P13, P16, P14, P20, P22, P26, P29, P30, P31, \\ P32, P33, P34, P35, P36, P38, P41, P45, P49\end{tabular} & 21 \\
 &Generic-fix/feature & \begin{tabular}[c]{@{}l@{}}P10, P11, P17, P24, P27, P37, P39, P40, P42, P46, P47\end{tabular} & 11 \\
 &Generic-self-interest & P3, P5, P9, P15, P21, P23 & 6 \\
 & Kinship/Altruism & P1, P4, P6, P12, P18, P19, P25, P28, P43, P44, P48 & 11 \\ \bottomrule
\end{tabular}
\end{table*}

\section{Quantitative Findings}
When asked about their preference to contribute, participants preferred to maintain their own libraries (i.e., 63\%).
Figure \ref{fig:demo_graph} also shows how participants split their time between maintaining their own libraries and contributing to another library in the ecosystem.
Regarding their contributions, it is unsurprising that participants spent most of their time maintaining a large number of their own libraries, rather than contributing to other libraries (i.e., more than two times per week, 39\% for maintaining their own and 8\% for contributing to others).
\begin{tcolorbox}
    \textbf{Finding 1} 
    Developers spend more time maintaining their code (i.e., 63\%), rather than contributing (6\%) to the ecosystem. 
\end{tcolorbox}
When making contributions to their own or other libraries, participants generally focused on stable libraries already in production (i.e., production, 31\% for maintaining their own and 46\% for contributing to others).
However, contributions to other libraries tended to focus on fixing bugs (26\%), then adding new features (20\%) and documentation (19\%), while participants tended to focus on adding new features (19\%) and maintenance activities such as refactoring, documentation, and writing test cases (18\%, 17\% and 16\% respectively) when maintaining their own libraries. 

Figure \ref{fig:R11} shows the quantitative answers to the survey questions regarding motivation, expertise drivers, and barriers.
In summary, we see that maintaining their own libraries is more motivated by personal drivers (38\%), while professional drivers are more responsible for contributions to other libraries (48\%).
Furthermore, the lack of time was agreed upon as the most cited barrier to maintaining and contributing to other libraries. 
\begin{tcolorbox}
    \textbf{Finding 2} 
    Maintaining their own libraries is driven by personal drivers (38\%), whereas contributions to the ecosystem are professionally driven (48\%).
\end{tcolorbox}

\section{Qualitative Findings}
\label{sec:findings}

We identified the following themes with their motivations and barriers, as shown in Table \ref{tab:my-table}.

\subsection{\textbf{Own-use-client - Developers who fix issues in libraries that they have explicitly stated they depend on.}}

Derived from the \textit{own-use} classification from Gerosa et al. \cite{Gerosa2021TheSS}, the \textit{own-use-client} motivation is the most frequent theme with 21 responses.
As explained by Gerosa et al. \cite{Gerosa2021TheSS}, this is an internalized
extrinsic motivation, where it is a self-regulating behavior based on external incentives and interventions.
In this case, developers contribute to other libraries because they want to fix issues in the library on which they depend.
This sentiment was also reflected in the quantitative responses, i.e., the second most frequent with 28\% in Figure \ref{fig:R11}.

An example of this theme is when there is something that is blocking or standing in the way of the developers' work:
\begin{quote}
\textit{``When they're broken and need fixing (and it's blocking my work), that's usually when I jump in.''} - Participant 2
\end{quote}

In this next example, the participant expresses the motivation to fix a bug explicitly in the library (packages) that developers depend on in their application. 
The example is shown below:

\begin{quote}
    \textit{``I sometimes fix a bug or do an improvement in packages I use''} - Participant 33
\end{quote}

Other reasons are more intrinsic to the maintainer to help with the maintenance of existing issues. 
An example is as follows:

\begin{quote}
\textit{``sometimes there are problems in other packages that i depend on, and the maintainers dont have the bandwidth to fix it.''} - Participant 31
\end{quote}

\subsection{\textbf{Generic-fix/feature - Developers who contribute by fixing or adding new features to libraries, but do not explicitly state their dependencies on them.}}
Similarly to the first theme, the \textit{generic-fix/feature} theme is based on the \textit{own-use} motivation.
Here, the focus is on explicitly fixing bugs in the other libraries where they are \textbf{not} a client (11 respondents). 
The motivation is less extrinsic, but to help other maintainers.
As shown in Figure \ref{fig:R11}, we find that the respondents considered finding a bug as the most frequent motivation to contribute to other libraries (i.e., 30\%), for example:

\begin{quote}
\textit{``Because I encounter some issue with them that needs to be fixed, and the easiest way is to fix it myself and contribute the changes back.''} - Participant 17
\end{quote}

Another example of this theme is when the participant expressed the need to generically fix any kind of bug that they encounter. 
In this case, the participant expresses the need to:

\begin{quote}
    \textit{``I do my best to fix bugs I come across''} - Participant 24
\end{quote}

Other responses were specific to fixing parts of the library or introducing new features to help meet their own needs:

\begin{quote}
\textit{``If they're broken in some way or I needed to add a feature to improve usability.''} - Participant 47
\end{quote}



\subsection{\textbf{Generic-self-interest - Developers who contribute for personal reasons, such as scratching an itch, wanting something for themselves, or for their own benefit.} }
The third theme is also related to generic extrinsic motivation, which is the common stereotype of open source developers.
For example, developers contribute by acting in their own interest.
Examples include:

\begin{quote}
\textit{Own itch. Something prevents me from proceeding and the change needs to be in their side. Even then, I normally try to work around instead of pushing a contribution.} - Participant 5
\end{quote}

Another example is when the developer feels the need to suggest a change to another library in the ecosystem.
This example is shown below:

\begin{quote}
\textit{``Usually because there are changes I want them to have''} - Participant 21
\end{quote}

In terms of barriers to contribution based on their self-interest, we also find that developers sometimes consider that much self-interest could be exercised by simply using forking:
\begin{quote}
\textit{``So much developer fork package and modified just for himself. Without pr issue or give an issue to package owner. He just did for himself, just for his business.''} - Participant 9
\end{quote}

Another example of a barrier is when the developer suffers from burnout due to the overwhelming duties of being a maintainer themselves.
For example:

\begin{quote}
\textit{``I'm extremely burned out due to maintaining a project with thousands of stars by myself for several years.''} - Participant 20
\end{quote}

\subsection{\textbf{Kinship/Altruism - Developers who contribute out of a sense of community belonging or responsibility.}}
Different from the other three themes, this theme is intrinsic and serves the nature of open source maintainers.
As defined, intrinsic motivation moves the person to act for the fun or challenge involved, which is different from the previous three themes.
Examples include:
\begin{quote}
    \textit{``Giving back to the community''}. - Participant 4
\end{quote}

\begin{quote}
    \textit{``The same reason I make my own packages: to better solve a problem for the entire ecosystem.''} - Participant 43
\end{quote}

In terms of barriers to kinship, one participant mentioned how the maturity of the open source project, might be a barrier for someone trying to make a legitimate contribution:

\begin{quote}
\textit{``The questionnaire made me think about why I contribute so little (in npm ecosystem). That’s likely due to the perceived (maybe imagined) burocracy of making a PR and fighting for it. Often times, it’s just lost energy (but this is based on corporate packages; not hobbyist). Also, if a repository doesn’t have proper testing, it makes me want to contribute \_less\_, since I feel I might break someone’s day...''} - Participant 5
\end{quote}

A similar sentiment was directed towards the ecosystem as a whole. In the following case, the participant expresses the magnitude of the ecosystem, where some parts might contain similar libraries.

\begin{quote}
\textit{``The node.js community is very hard to work with. People keep reinventing the same thing over and over and it confused on boarding new people''} - Participant 15
\end{quote}

\begin{tcolorbox}
    \textbf{Finding 3:} 
     We identify four themes that motivate developers when contributing to the ecosystem:
    (i.e., own-use-client, generic-fix/feature, generic-self-interest, and kinship/altruism).
    The most prevalent motivation to contribute is to address issues in libraries that developers depend on (i.e., own-use-client), while a large ecosystem is seen as a potential barrier for maintainers to contribute back.
\end{tcolorbox}

\section{Threats to Validity}
\label{sec:validity}
The first threat is related to survey participants misunderstanding survey questions.
To mitigate this threat, we conducted a pilot survey to confirm the understanding of the participants. 
Furthermore, we decided to use existing work to cover all motivations (Subramanian et al. \cite{Subramanian:IEEE2022} and Vadlamani and Baysal \cite{Vadlamani:icsme2020}).
The final threat is related to the
generalizability of our results to other library ecosystems, which we cannot claim.
We plan to expand our study to other ecosystems (PyPI for Python and Maven for Java) in the future to mitigate this threat.

\epigraph{\textit{\textbf{Q. Why do you contribute to other npm libraries, i.e., libraries which you do not own?}} \\
\textit{``A. We live in a society''}}{Participant 28}

\section{Conclusion and Future Directions}
\label{sec:implications}
As a new idea paper, we now indicate key implications and future work.

\textbf{Similar to how motivations of individuals in a society can change based on their role, the motivations of contributors in an ecosystem can also change depending on the role they play.}
Finding 3 suggests that a client is more likely to fix a library they depend on, but also hints at the struggles that maintainers face when making contributions.
Finding 2 also shows that different motivations arise depending on the role that the developer plays, either as the maintainer or a user of that library.
A better understanding of this relationship strengthens our claim that we live in a society that requires mutual contributions between members of the ecosystem.

\textbf{Using clients as potential contributors to the libraries on which they depend.}
In the context of an OSS ecosystem, Finding 1 and Finding 2 illustrate that there is a difference in motivation from when they decide to contribute to the ecosystem. 
These contributions could have the potential to fix important bugs and implement new features that may be needed by the library.

\textbf{Future Plans.}
Motivated by these results, there are many possibilities for future research.
One important avenue is to conduct more empirical research to better understand contributions to a library by its clients, thereby allowing us to differentiate these contributions from those made by others in the ecosystem. Not all clients are the same -- understanding different client-library relationships and their implications is another important research direction.

From a human perspective, there are also opportunities to study the interactions and communication between different roles in the ecosystem, such as the discussions that occur during code review between clients and maintainers. Additionally, exploring the motivations and incentives for clients to contribute to the libraries they depend on could provide valuable insights for encouraging more participation from this group.





\bibliographystyle{IEEEtran}
\bibliography{filtered_ref}

\end{document}